\documentclass[preprint,aps,amsmath,nofootinbib,tightenlines,superscriptaddress]{revtex4}
\usepackage{bm}
\usepackage{epsfig}
\usepackage[dvips]{hyperref} 

\newif\ifpdf
\ifx\pdfoutput\undefined
\pdffalse 
\else
\pdfoutput=1 
\pdftrue
\fi

\def\nslash{n\!\!\!\slash}
\def\bnslash{\bar n\!\!\!\slash}

\def\OMIT#1{}

\newcommand{\nn}{\nonumber} 

\newcommand{\bn}{{\bar n}}

\newcommand{\bnP}{\bar {\cal P}}

\newcommand{\mcdot}{\!\cdot\!}

\newcommand{\SCETa}{\mbox{${\rm SCET}_{\rm I}$ }}
\newcommand{\SCETb}{\mbox{${\rm SCET}_{\rm II}$ }}

\begin{document}
\ifpdf
\DeclareGraphicsExtensions{.pdf, .jpg}
\else
\DeclareGraphicsExtensions{.eps, .jpg}
\fi


\preprint{ \vbox{\hbox{MIT-CTP-3549} 
}}

\title{\phantom{x}
\vspace{0.5cm}
Heavy Quark Symmetry in Isosinglet
 Nonleptonic $B$-Decays \\ 
\vspace{0.6cm}
}

\author{Andrew E. Blechman}
\affiliation{Department of Physics and Astronomy, The Johns Hopkins University,\\
        Baltimore, MD 21218\footnote{Electronic address: blechman@pha.jhu.edu}
\vspace{0.4cm}}
\author{Sonny Mantry}
\affiliation{Center for Theoretical Physics, Massachusetts Institute for
Technology,\\ Cambridge, MA 02139\footnote{Electronic address: mantry@mit.edu, 
iains@mit.edu}
\vspace{0.2cm}}
\author{Iain W. Stewart\vspace{0.4cm}}
\affiliation{Center for Theoretical Physics, Massachusetts Institute for
Technology,\\ Cambridge, MA 02139\footnote{Electronic address: mantry@mit.edu, 
iains@mit.edu}
\vspace{0.2cm}}

\vspace{0.5cm}

\begin{abstract}
\vspace{0.3cm}

We use a factorization theorem from the soft-collinear effective theory along
with heavy quark symmetry to make model independent predictions for $\bar B^0\to
D^{(*)0}M$ where $M=\{\eta,\eta',\phi, \omega\}$. Gluon production of these
isosinglet mesons is included. We predict the equality of branching fractions in
the $\bar B\to DM$ and $\bar B\to D^*M$ channels, with corrections at order
$\Lambda _{\rm QCD}/Q$ and $\alpha_s(Q)$ where $Q=m_b,m_c$, or $E_M$.  We also
predict that $Br(\bar B^0 \to D\eta')/Br(\bar B^0\to D\eta)=\tan^2\theta = 0.67$
and $Br(\bar B\to D\phi)/Br(\bar B\to D \omega)\lesssim 0.2$, where here there
are also $\alpha_s(\sqrt{E\Lambda})$ corrections.  These results agree well with
the available data.  A test for SU(3) violation in these decays is constructed.

\end{abstract}

\maketitle


Nonleptonic weak-decays involving $b\to c \bar q q'$ transitions provide an
interesting framework for testing power expansions and factorization in QCD at
the $m_b\sim 5\,{\rm GeV}$ scale.  The Soft-Collinear Effective Theory
(SCET)~\cite{Bauer:2000ew,Bauer:2000yr,Bauer:2001ct,Bauer:2001yt} has been used
to make predictions for two-body non-leptonic $b\to c$ decays such as color
allowed decays $\bar B\to D^{(*)}M^-$ where $M={\pi, \rho, K,
  K^*}$~\cite{Bauer:2001cu}, color suppressed decays $\bar B^0\to
D^{(*)0}M^0$~\cite{Mantry:2003uz}, decays to excited $D$ mesons $\bar B\to
D^{**} M$~\cite{Mantry:2004pg}, as well as baryon decays $\Lambda_b\to \Lambda_c
M$ and $\Lambda_b\to \Sigma_c^{(*)}M$~\cite{Leibovich:2003tw}.  These
predictions make use of a systematic expansion in $\Lambda_{\rm QCD}/m_{b,c}$
and $\Lambda_{\rm QCD}/E_M$.  For earlier work on color allowed decays
see~\cite{Dugan:1990de,Politzer:1991au,Blok:1992na,Beneke:2000ry}. The nature of
factorization has also been studied in inclusive $B\to D^{(*)}X$ decays, as well
as decays to multi-body final states like $B\to D\pi\pi\pi\pi$, and decays to
higher spin
mesons~\cite{Bauer:2002sh,Ligeti:2001dk,Balzereit:1998ei,Diehl:2001xe}.

The Belle and BaBar Collaborations have recently reported measurements of the
color suppressed decay channels $\bar B^0\to D^{(*)0}\eta$, $\bar B^0\to
D^{0}\eta'$, and $\bar B^0\to D^{(*)0}\omega $ which have an isosinglet meson
$M$ in the final state~\cite{Aubert:2003sw,Abe:2004sq,Aubert:2004bf}. A
summary of the data is given in Table I. By now it is well understood that
``naive'' factorization~\cite{Wirbel:1985ji} fails miserably for these
``color-suppressed'' decays.  A rigorous framework for discussing them in QCD is
provided by the factorization theorem derived in Ref.~\cite{Mantry:2003uz}.  The
presence of isosinglet mesons enriches the structure of the decays due to
$\eta$--$\eta'$ and $\omega$--$\phi$ mixing effects and gluon production
mechanisms~\cite{Feldmann:1999uf,Kroll:2002nt,Beneke:2002jn}.  In this paper, we
generalize the SCET analysis of~\cite{Mantry:2003uz} to include isosinglets. We
also construct a test of SU(3) flavor symmetry in color suppressed decays, using
our results to include the $\eta-\eta'$ mixing.

\begin{table}[b!]
\begin{center}
\begin{tabular}{|c|c|c|c|c|c|}
\hline
Decay & Br$(10^{-4})$ (BaBar) & Br$(10^{-4})$ (Belle) &
Br$(10^{-4})$ (Avg.) & $|A|$ ($10^{-4}$ MeV)  \\
\hline\hline
$\bar B^0 \to D^0 \eta$ & $2.5\pm 0.2\pm 0.3$  & $1.83\pm 0.15\pm0.27$
 & $2.1\pm 0.2$ & $1.67\pm 0.09$  \\
$\bar B^0 \to D^{*0} \eta$ & $2.6\pm 0.4\pm 0.4$ & $-$
 & $2.6\pm 0.6$ & $1.87\pm 0.22$  \\
$\bar B^0 \to D^0 \eta'$ &  $1.7\pm 0.4 \pm 0.2$ & $1.14\pm 0.20\pm 0.11$
 & $1.3\pm 0.2$ & $1.31\pm 0.11$  \\
$\bar B^0 \to D^{*0} \eta'$ & $1.3\pm 0.7 \pm 0.2$ & $1.26\pm 0.35\pm 0.25$  
 & $1.3\pm 0.4$ & $1.33\pm 0.19$  \\
$\bar B^0 \to D^0 \omega$ & $3.0\pm 0.3 \pm 0.4$  & $2.25\pm 0.21\pm 0.28$
 & $2.5\pm 0.3$ & $1.83\pm 0.11$  \\
$\bar B^0 \to D^{*0} \omega$ & $4.2 \pm 0.7 \pm 0.9$ & $-$
 & $4.2 \pm 1.1$ & $2.40\pm 0.31$  \\
$\bar B^0 \to D^{(*)0} \phi$ & $-$ & $-$
 & $-$ & $-$  \\
\hline
$\bar B^0 \to D^{0} \pi^0$ & $2.9\pm 0.2\pm 0.3$ & $2.31\pm 0.12\pm 0.23$
 & $2.5 \pm 0.2$ & $1.81 \pm 0.08$  \\ 
$\bar B^0 \to D^{*0} \pi^0$ & $-$ & $-$
 & $2.8 \pm 0.5$ & $1.95 \pm 0.18$  \\ 
$\bar B^0 \to D^{0} \bar K^0$ & $0.62 \pm 0.12 \pm 0.04$ 
                              & $0.50^{+0.13}_{-0.12}\pm 0.06$
 & $0.44 \pm 0.06$ & $0.76\pm 0.06$  \\ 
$\bar B^0 \to D^{*0} \bar K^0$ & $0.45 \pm 0.19 \pm 0.05$  & $<0.66$
 & $0.36 \pm 0.10$ & $0.69\pm 0.10$  \\ 
$\bar B^0 \to D^{+}_s K^-$ & $0.32 \pm 0.10 \pm 0.10$ 
                           & $0.293\pm 0.055 \pm 0.079$
 & $0.30 \pm 0.08$ & $0.64\pm 0.08$  \\ 
\hline
\end{tabular}
\end{center}
{\caption{Data on $B\to D$ and $B\to D^*$ decays with isosinglet light mesons
    and the weighted average. The BaBar data is from Ref.~\cite{Aubert:2003sw} 
   and the Belle data is from Refs.~\cite{Abe:2004sq} 
  and~\cite{Drutskoy:2004}. }
\label{table_data} }
\end{table}

\begin{figure}[!t]
\vskip0.1cm
 \centerline{
  \mbox{\epsfxsize=4.5truecm \hbox{\epsfbox{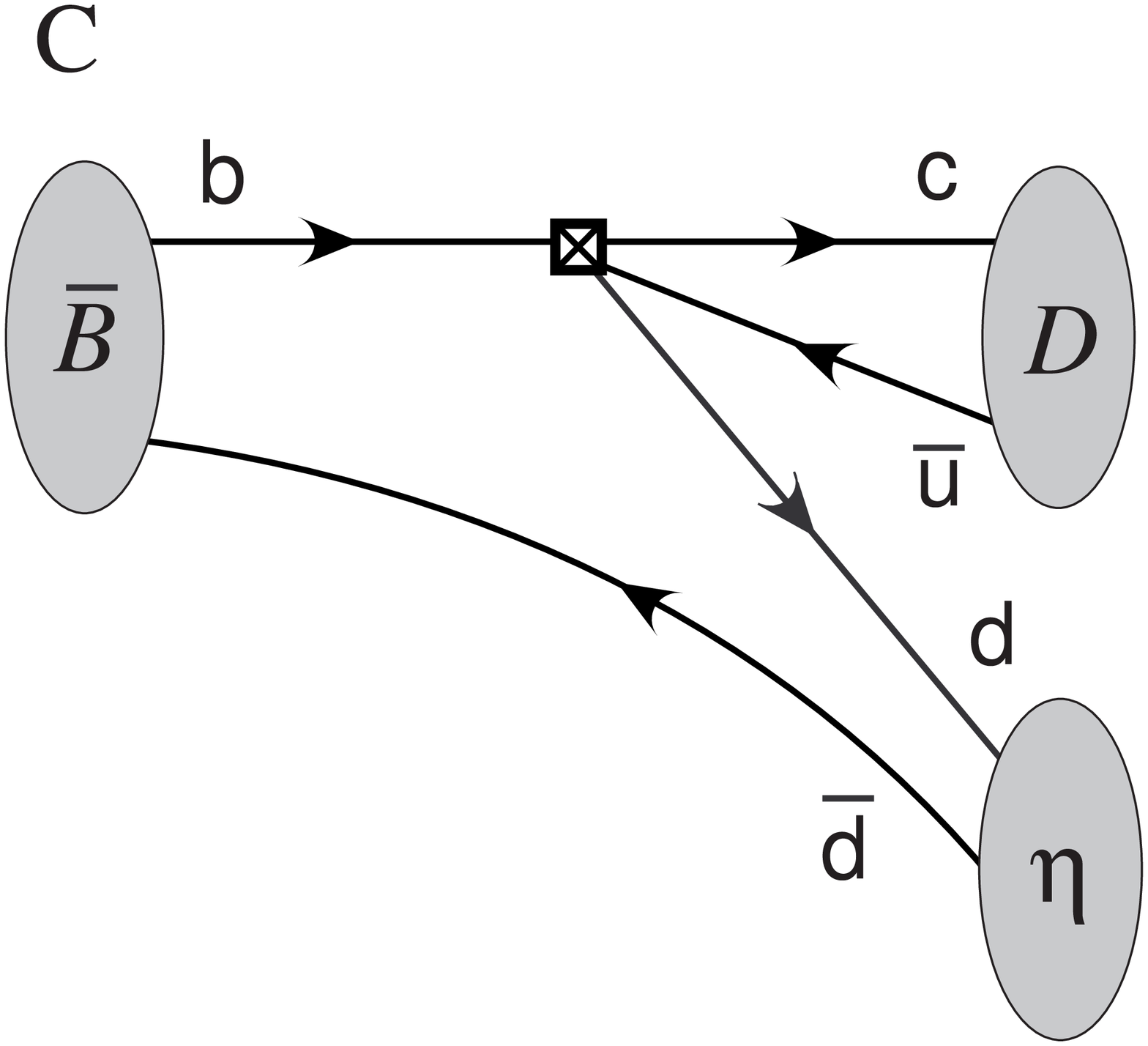}} }
   \hspace{0.7cm}
  \mbox{\epsfxsize=4.5truecm \hbox{\epsfbox{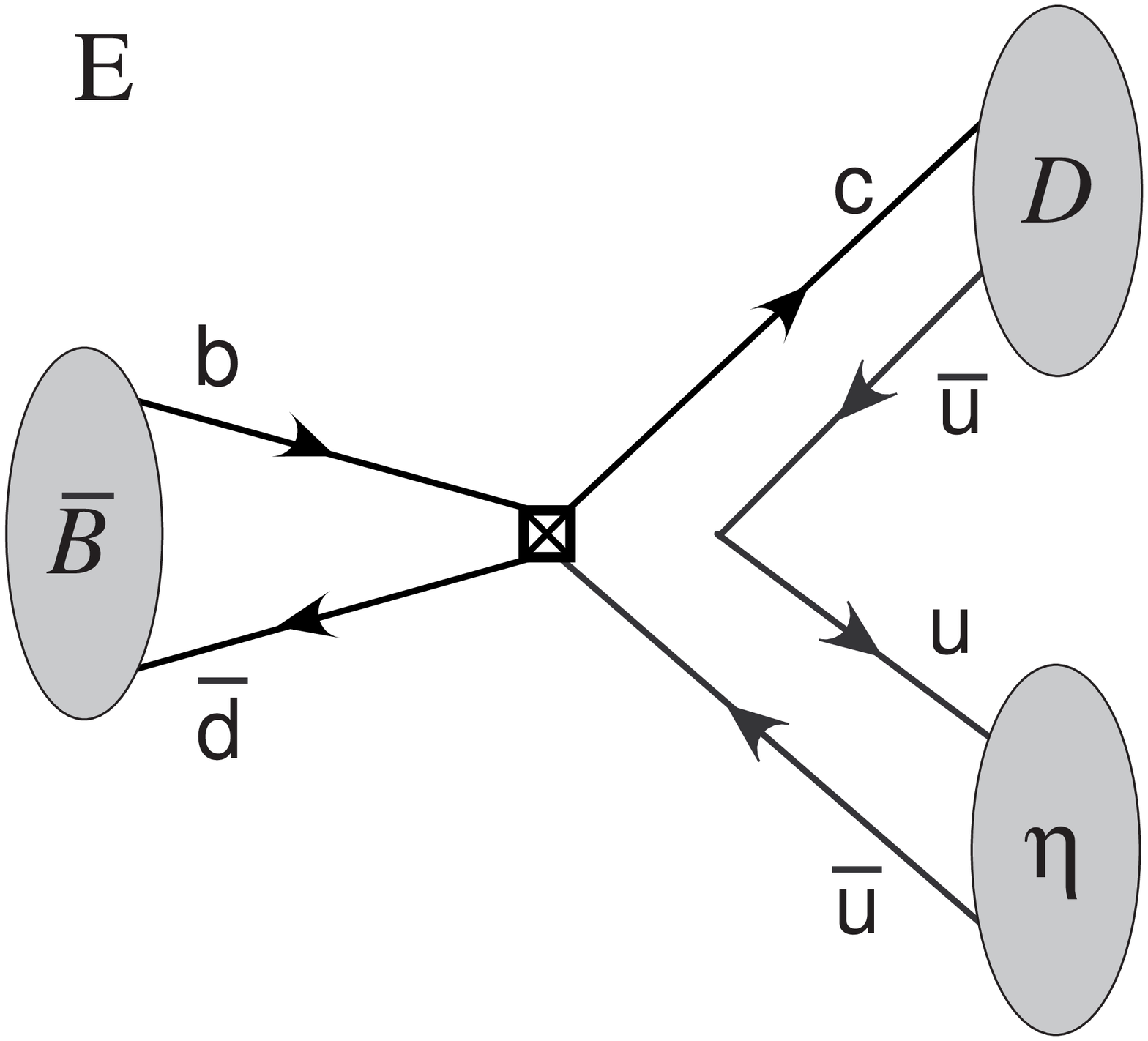}} }
   \hspace{0.7cm}
  \raisebox{0.2cm}{\epsfxsize=4.5truecm \hbox{\epsfbox{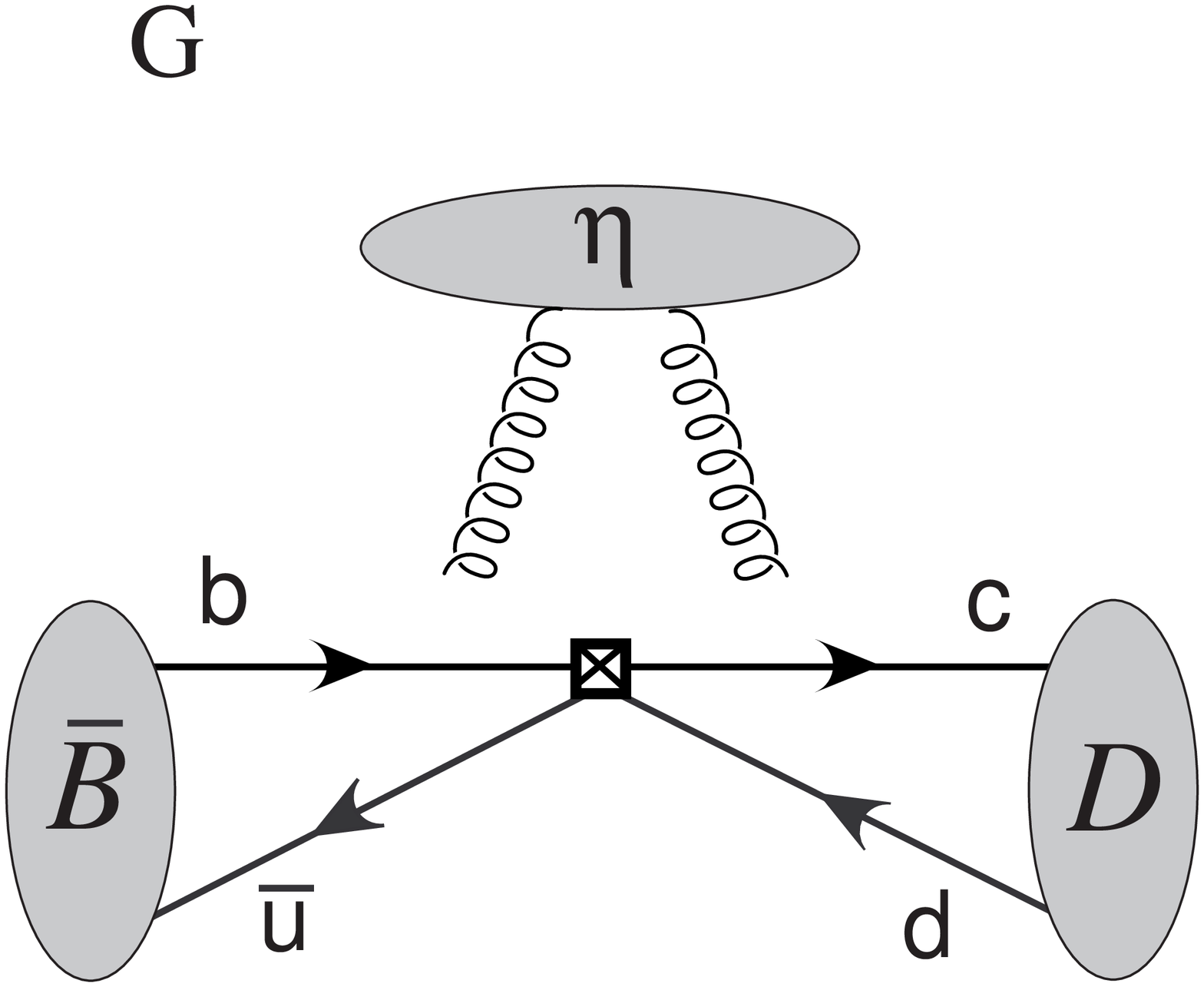}} }
  } 
 \vspace{0.2cm}
\vskip-0.3cm
\caption[1]{Flavor diagrams for $\bar B\to D\eta$ decays, referred to 
  as color-suppressed (C), $W$-exchange (E), and gluon production
  (G). These amplitudes denote classes of Feynman diagrams where the remaining
  terms in a class are generated by adding any number of gluons as well as
  light-quark loops to the pictures.}
\label{fig_qcd} 
\vskip0cm
\end{figure}

The quark level weak Hamiltonian is
\begin{eqnarray}\label{Hw}
 {\cal H}_W = \frac{G_F}{\sqrt2} V_{cb} V_{ud}^* [ C_1(\mu)
 (\bar c b)_{V-A} (\bar d u)_{V-A} + C_2(\mu)
 (\bar c_i b_j)_{V-A} (\bar d_j u_i)_{V-A} ]\,,
\end{eqnarray}
where $C_1$ and $C_2$ are Wilson coefficients.  For color-suppressed decay
channels it gives rise to three flavor amplitudes denoted $C$, $E$, and $G$ in
Fig.~\ref{fig_qcd}, which take on a precise meaning in terms of operators in the
SCET analysis at leading order in $\Lambda_{QCD}/Q$. Here $Q$ is a hard scale
on the order of the heavy quark masses $m_b, m_c$ or the isosinglet meson energy
$E_{M}$. The gluon $G$ amplitude is unique to isosinglet mesons. We will show
however that for $B\to D^{(*)}M$ decays the $G$ amplitude is suppressed by
$\alpha_s(\sqrt{E\Lambda})$ relative to the $C,E$ contributions.

For color suppressed decays to isosinglet mesons $M=\{\eta,\eta',\omega,\phi\}$
we will show that the factorization theorem for the amplitudes
$A^{(*)}_{M}=\langle D^{(*)0} M | H_W | \bar B^0 \rangle$ is
\begin{eqnarray}\label{result}
 A^{(*)M} &=& A_{\rm short}^{(*)M} + A_{\rm glue}^{(*)M}  
  + A_{\rm long}^{(*)M}\ \pm (L\leftrightarrow R) \,,
\end{eqnarray}
where the $\pm$ refers to the cases $DM$, $D^*M$ and the three amplitudes at LO are
\begin{eqnarray} \label{ftheorem}
  A_{\rm short}^{(*)M} &=& N_q^M \sum_{i=0,8}
     \int_0^1\!\!\!dx\, dz\!\!  \int\!\! dk_1^+ dk_2^+\, 
   C^{(i)}_{L}(z)\: 
   J^{(i)}_q(z,x,k_1^+,k_2^+)\: S^{(i)}_L(k_1^+,k_2^+)\:  \phi^M_q(x) \,, \\
%
A_{\rm glue}^{(*)M} &=& N_g^M \sum_{i=0,8}
     \int_0^1\!\!\!dx\, dz\!\!  \int\!\! dk_1^+ dk_2^+\, 
   C^{(i)}_{L}(z)\: 
   J^{(i)}_g(z,x,k_1^+,k_2^+)\: S^{(i)}_L(k_1^+,k_2^+)\:  \overline \phi^M_g(x) \,, \nn \\
%
  A_{\rm long}^{(*)\!M} &=& N_q^M \sum_{i=0,8}
   \int_0^1\!\!\! dz\!\!  \int\!\! dk^+ d\omega \!\! \int\!\! d^2\!x_\perp 
   C^{(i)}_{L}(z)\: 
   \bar J^{(i)}(\omega k^+)\: \Phi_{L}^{(i)}(k^+\!,x_\perp,\varepsilon^*_{D^*}) 
    \Psi_M^{(i)}(z,\omega,x_\perp,\varepsilon^*_M) \nn ,
\end{eqnarray}
where $i=0,8$ are for two different color structures.  Here $A_{\rm
  short}^{(*)M}$ and $A_{\rm long}^{(*)\!M}$ are very similar to the results
derived for non-singlet mesons in Ref.~\cite{Mantry:2003uz}, and each contains a
flavor-singlet subset of the sum of $C$ and $E$ graphs. The amplitude $A_{\rm
  glue}^{(*)M}$ contains the additional gluon contributions. The $S_L^{(0,8)}$
are universal generalized distribution functions for the $B\to D^{(*)}$
transition.  The $\phi_{q,g}^M$ are leading twist meson distribution functions,
and~\footnote{For Cabbibo suppressed channels we replace $V_{ud}^*\to V_{us}^*$
  in $N_q^M$ and $N_g^M$.}
\begin{eqnarray} \label{Nqg}
  N_q^M =  \frac{1}{2} \, f^M_q  G_F V_{cb}^{\phantom{*}} V_{ud}^*   
  \:  \sqrt{m_B m_{D^{(*)}}}\,,
  \qquad
  N_g^M =  \sqrt{\frac{8}{3}}\, f_1^M \,  G_F V_{cb}^{\phantom{*}} V_{ud}^* 
  \:   \sqrt{m_B m_{D^{(*)}}} \,.
\end{eqnarray}
The $\Phi_L^{(i)}$ and $\Psi_M^{(i)}$ are long distance analogs of $S_L^{(i)}$
and $\phi^M$ where the $x_\perp$ dependence does not factorize.  At lowest order
in the perturbative expansion, $C_L^{(0)}=C_1+C_2/3$ and $C_L^{(8)}=2 C_2$ and
are independent of the parameter $z$. The $(L\leftrightarrow R)$ terms in
Eq.~(\ref{result}) have small coefficients $C_R^{(0,8)}\sim {\cal
  O}(\alpha_s(Q))$ and will be neglected in our phenomenological analysis.
Finally, the jet functions $J_q^{(i)}$, $J_g^{(i)}$, and $\bar J^{(i)}$ are
responsible for rearranging the quarks in the decay process; they can be
computed in perturbation theory and are discussed further below.

\begin{figure}[!t]
\vskip-0.3cm
 \centerline{ 
  \vbox{\raisebox{-0.15cm}{SCET${}_{\rm I}$}\hspace{3.4cm}
        \raisebox{-0.15cm}{SCET${}_{\rm II}$ (long)}\hspace{3.cm}
        \raisebox{-0.15cm}{SCET${}_{\rm II}$ (short)} \hspace{-1.5cm}
 } }
 \centerline{\raisebox{1.5cm}{C} \hspace{0.1cm}
  \mbox{\epsfxsize=5.1truecm \hbox{\epsfbox{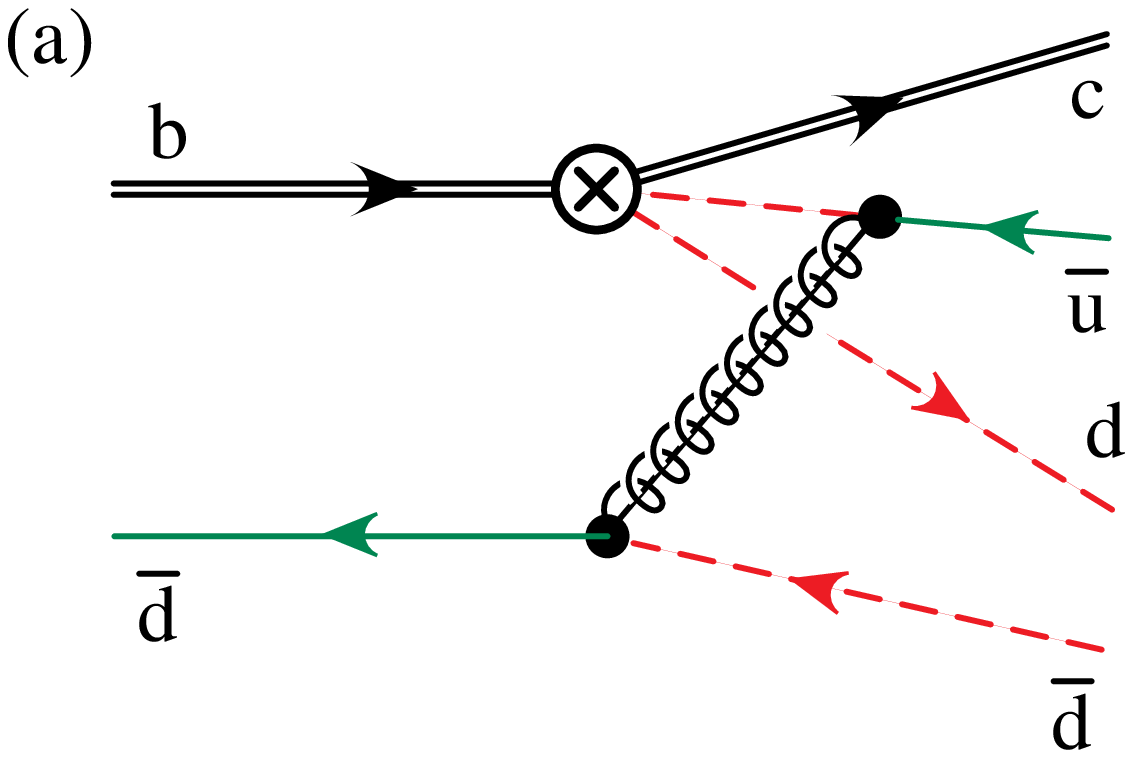}} }
  \hspace{0.3cm}
  \raisebox{0.2cm}
   {\mbox{\epsfxsize=5.0truecm \hbox{\epsfbox{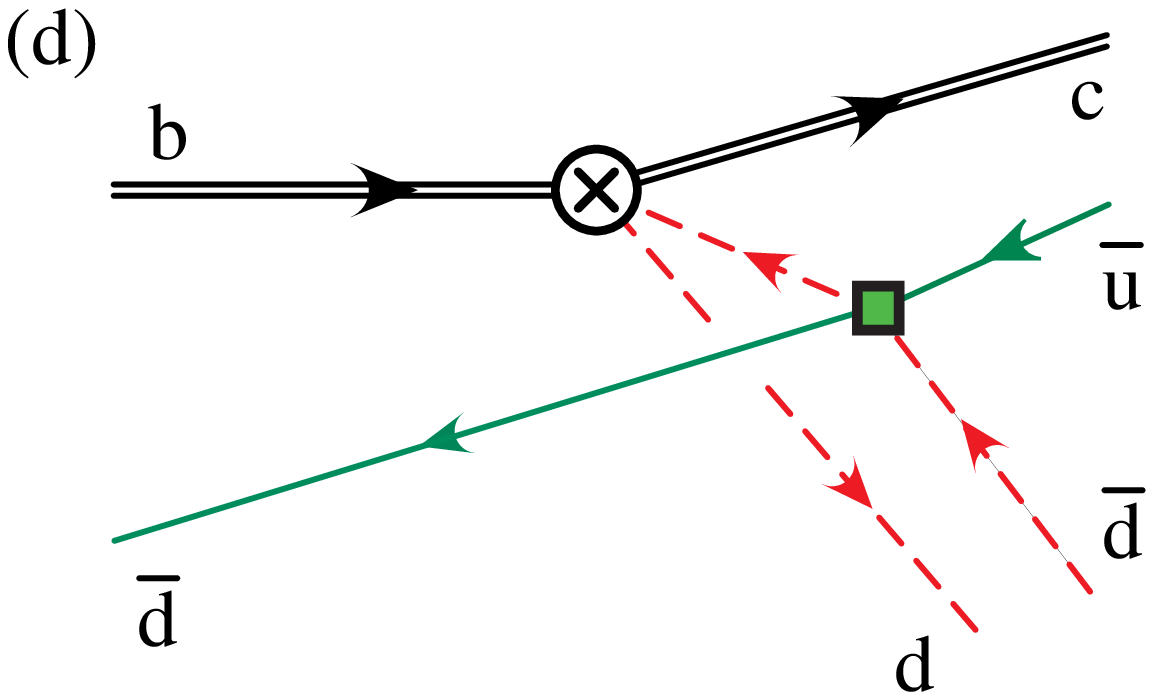}}} }
  \hspace{-0.cm}
  \raisebox{-0.cm}
   {\mbox{\epsfxsize=4.8truecm \hbox{\epsfbox{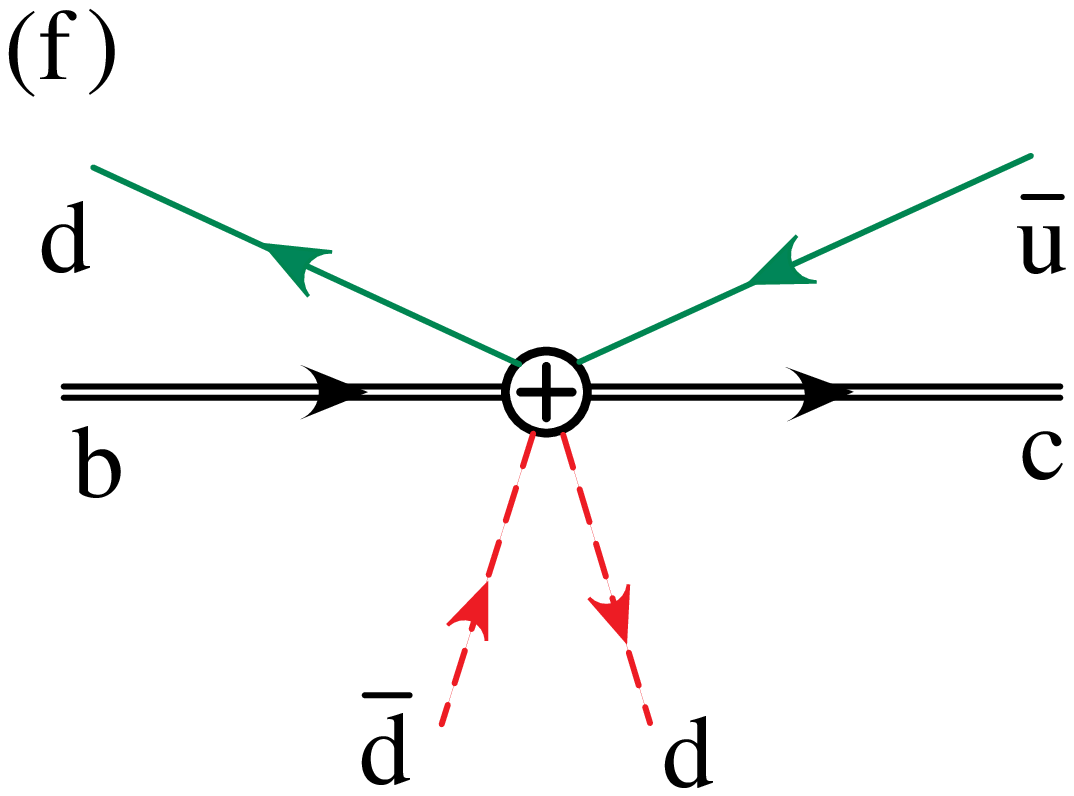}}} }
  }
\vskip-0.6cm
  \centerline{\raisebox{1.5cm}{E} \hspace{0.1cm}
  \mbox{\epsfxsize=5.3truecm \hbox{\epsfbox{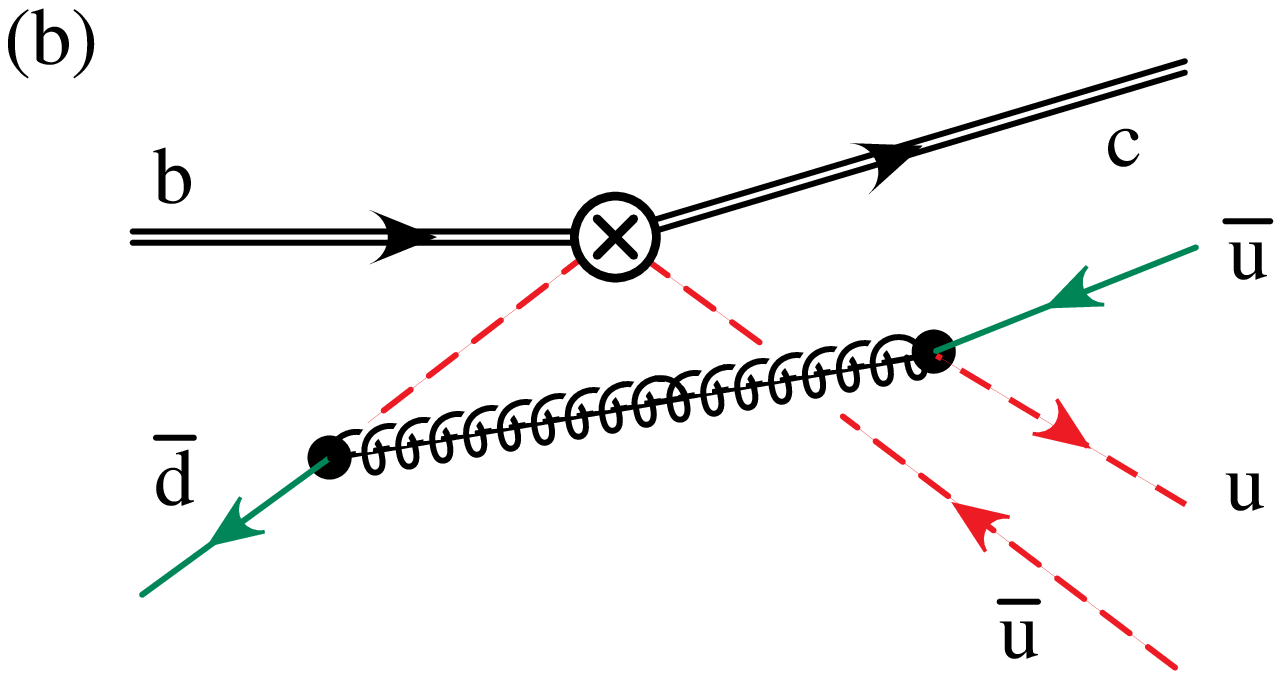}} }
  \hspace{0.cm}
  \raisebox{-0.2cm}
  {\mbox{\epsfxsize=5.1truecm \hbox{\epsfbox{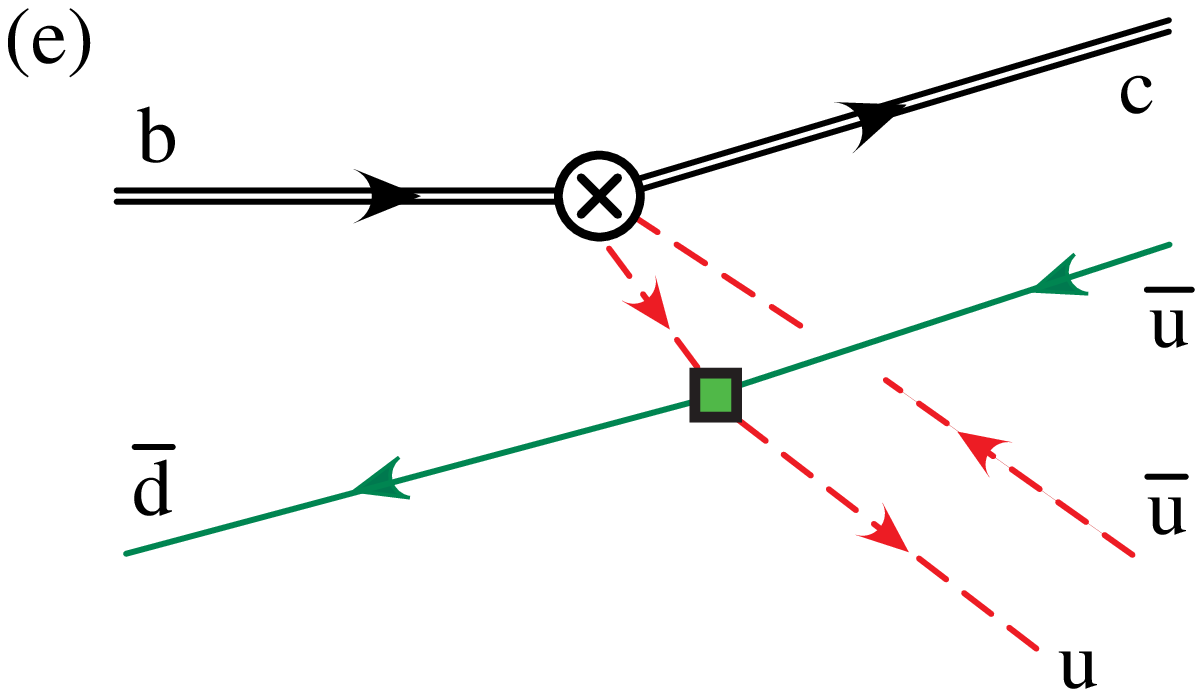}}} }
   \hspace{-0.cm}
  \raisebox{-0.2cm}
   {\mbox{\epsfxsize=4.8truecm \hbox{\epsfbox{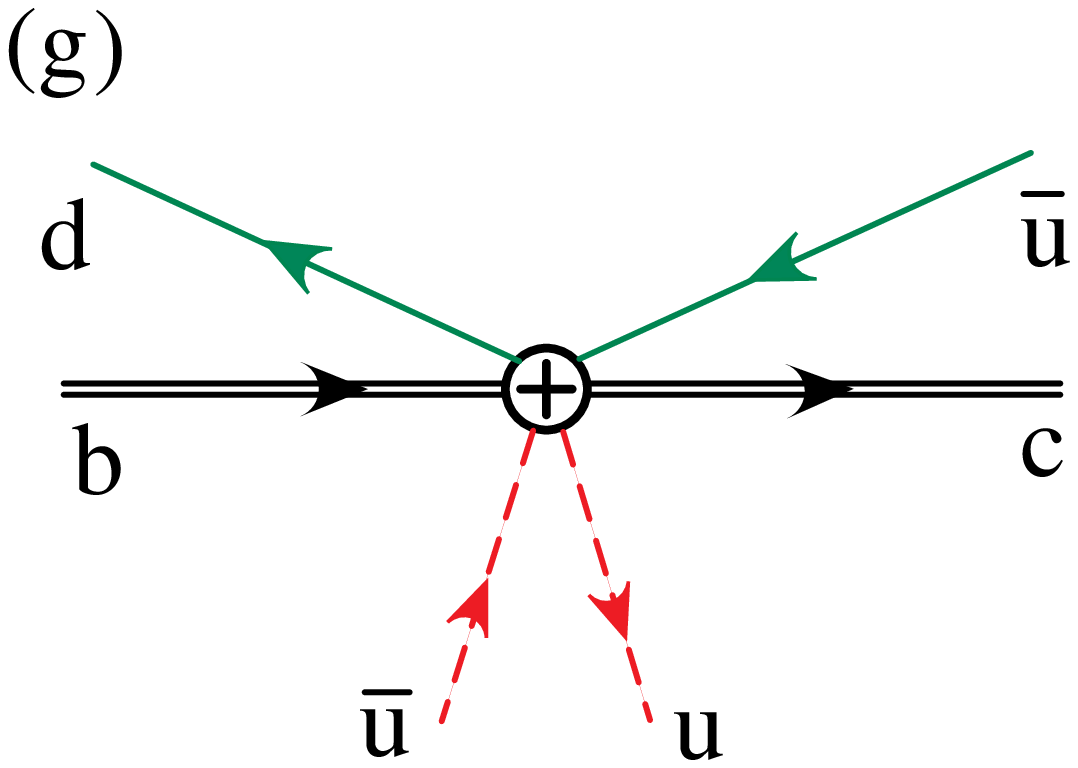}}} }
  }
\vskip-0.6cm
  \centerline{\raisebox{1.5cm}{G} \hspace{0.1cm}
  \raisebox{0.3cm}\
  {\mbox{\epsfxsize=5.1truecm \hbox{\epsfbox{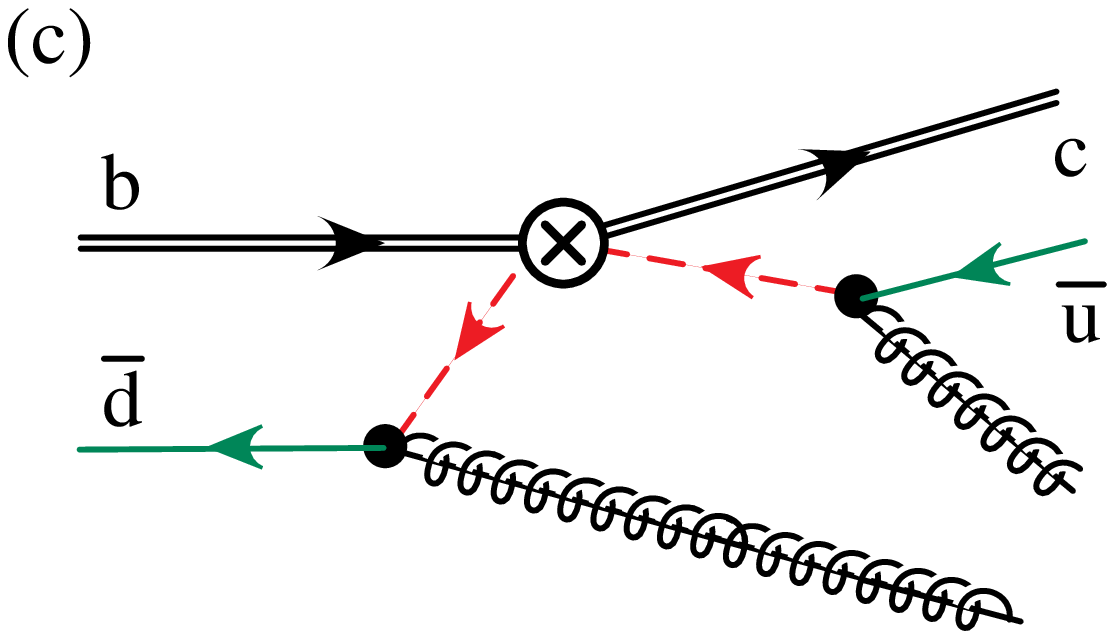}}} }
  \phantom{x}\hspace{5.2cm}
  \raisebox{0.3cm}
   {\mbox{\epsfxsize=4.9truecm \hbox{\epsfbox{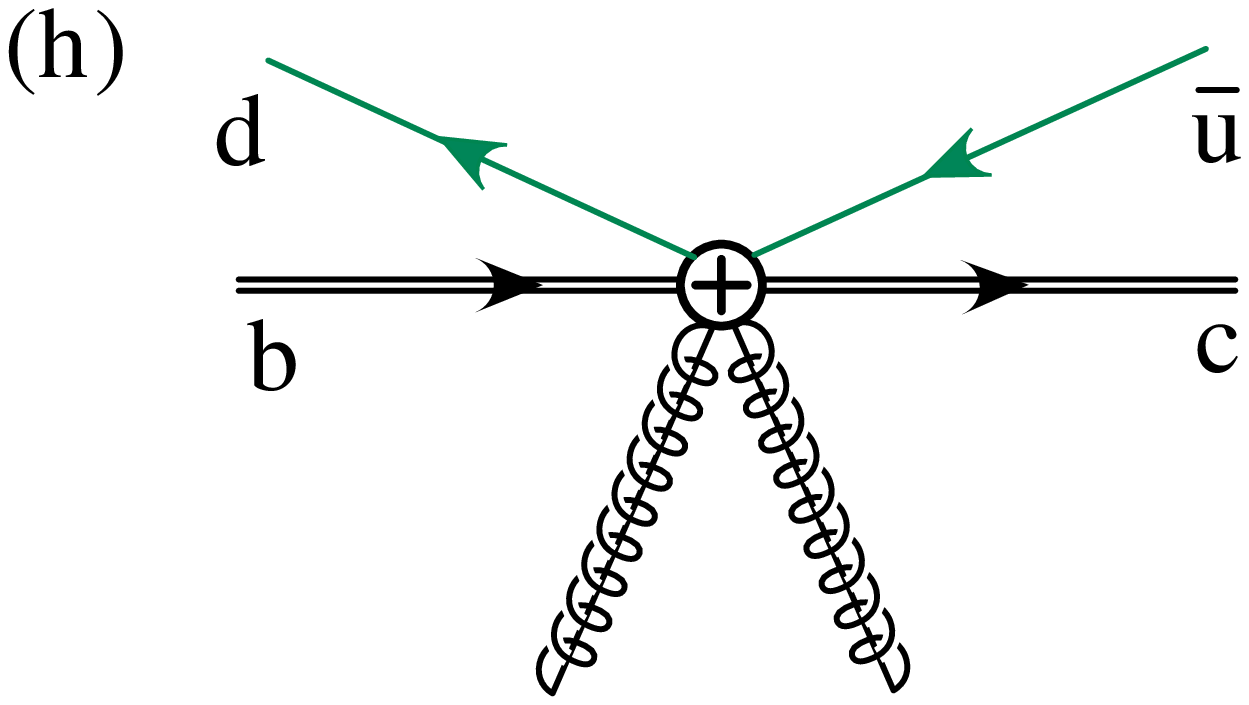}}} }
  }
\vskip-0.3cm
\caption[1]{Graphs for the tree level matching calculation from
  \SCETa (a,b,c) onto \SCETb (d,e,f,g,h). The dashed lines are collinear quark
  propagators and the spring with a line is a collinear gluon. Solid lines are
  quarks with momenta $p^\mu\sim \Lambda$. The $\otimes$ denotes an insertion of
  the weak operator in the appropriate theory.  The solid dots in (a,b,c) denote
  insertions of the mixed usoft-collinear quark action ${\cal L}_{\xi q}^{(1)}$.
  The boxes in (d,e) denote the \SCETb operator ${\cal L}_{\xi\xi qq}^{(1)}$
  from Ref.~\cite{Mantry:2003uz}.}
\label{fig_scet1} 
\vskip0cm
\end{figure}

The derivation of Eq.~(\ref{ftheorem}) involves subsequently integrating out the
scales $Q=\{m_b,m_c,E_M\}$ and then $\sqrt{E_M\Lambda_{\rm QCD}}$ by matching
onto effective field theories, ${\rm QCD}\to \SCETa\to \SCETb$, and we refer to
Ref.~\cite{Mantry:2003uz} for notation and further details. Here we only give
the reader a sense of the procedure, and discuss additions needed for the
isosinglet case.  In \SCETa there is only a single time-ordered product for
color suppressed decays
\begin{eqnarray}\label{Tprodorig}
  T_{L,R}^{(0,8)} &=& \frac12 
  \int\!\! d^4\!x\, d^4\!y\ T\big\{ {\cal Q}_{L,R}^{(0,8)}(0)\,, 
   i{\cal L}_{\xi q}^{(1)}(x)\,, i{\cal L}_{\xi q}^{(1)}(y)\big\}\,.
\end{eqnarray}
Here ${\cal Q}_{L,R}^{(0,8)}(0)$ are the LO operators in \SCETa that $H_W$ gets
matched onto, and ${\cal L}_{\xi q}^{(1)}$ is the subleading ultrasoft-collinear
interaction Lagrangian, which is the lowest order term that can change a
ultrasoft quark into a collinear quark. The power suppression from the two
${\cal L}_{\xi q}^{(1)}$'s makes the amplitudes for color suppressed decays
smaller by $\Lambda/Q$ from those for color allowed decays.  The $C$, $E$, and
$G$ diagrams in Fig.~\ref{fig_qcd} are different contractions of the terms in
$T_{L,R}^{(0,8)}$, and at tree level are given by Figs.~\ref{fig_scet1}(a),
\ref{fig_scet1}(b), and \ref{fig_scet1}(c) respectively. The propagators in
these figures are offshell by $p^2\sim E_M\Lambda$. In \SCETb all lines are
offshell by $\sim \Lambda^2$, so the propagators either collapse to a point as
shown in Figs.~\ref{fig_scet1}(f), \ref{fig_scet1}(g), and \ref{fig_scet1}(h),
or the quark propagator remains long distance as denoted in
Figs.~\ref{fig_scet1}(d) and \ref{fig_scet1}(e). For the terms in the
factorization theorem in Eq.~(\ref{ftheorem}), Figs.~\ref{fig_scet1}(f,g)
contribute to $A_{\rm short}$, Fig.~\ref{fig_scet1}(h) contributes to $A_{\rm
  glue}$, and Figs.~\ref{fig_scet1}(d,e) contributes to $A_{\rm long}$. A
notable feature is the absence of a long distance gluon contribution. Momentum
conservation at the ${\cal L}_{\xi q}^{(1)}$ vertex forbids the quark
propagators in Fig.~\ref{fig_scet1}(c) from having a long distance component (or
more generally there does not exist an appropriate analog of the shaded box
operator in Figs.~\ref{fig_scet1}(d,e) that takes a soft $\bar d$ to a soft
$\bar u$).

\OMIT{The \SCETa diagrams are next matched onto \SCETb to give the 
operators $C_{II}^{long}$, $E_{II}^{long}$, $CE_{II}^{short}$ and 
$G_{II}^{short}$ in Fig.~\ref{fig_scet2}. In 
$C_I$ and $E_I$, momentum conservation in the $p^+$ component
forbids the gluon to propagate in \SCETb and must be integrated out. On the
other hand, the collinear quark propagator can have off-shellness $p^2\sim
E\Lambda _{QCD}$(hard-collinear) or $p^2\sim \Lambda _{QCD}^2$(collinear)
giving rise to the short distance($CE_{II}^{short}$) or long
distance($C_{II}^{long}, E_{II}^{long}$) operators respectively~\cite{mps}. The
$G_I$ diagram in \SCETa gives rise to the additional short distance operator
$G_{II}^{short}$ in \SCETb. Momentum conservation again forbids a corresponding long
distance contribution from $G_I$.
}

The diagrams in Fig.~\ref{fig_scet1}(f,g) have isosinglet and isotriplet
components. The corresponding isosinglet operators in \SCETb
are~\cite{Mantry:2003uz}
\begin{eqnarray} \label{OV2}
 O_{j}^{(0)}(k^+_i,\omega_k) &=& 
  \Big[ \bar h_{v'}^{(c)}  \Gamma^h_j\,  h_v^{(b)} \:
  (\bar d\,S)_{k^+_1} \nslash P_L\, (S^\dagger u)_{k^+_2} \Big]
  \Big[ (\bar \xi_n^{(q)} W)_{\omega_1} \Gamma_c 
  (W^\dagger \xi_n^{(q)})_{\omega_2} \Big]\,, \\
 O_{j}^{(8)}(k^+_i,\omega_k) &=& 
  \Big[ (\bar h_{v'}^{(c)} S) \Gamma^h_j\, T^a\, (S^\dagger h_v^{(b)}) \:
  (\bar d\,S)_{k^+_1} \nslash P_L T^a (S^\dagger u)_{k^+_2} \Big]
  \Big[ (\bar \xi_n^{(q)} W)_{\omega_1} \Gamma_c 
  (W^\dagger \xi_n^{(q)})_{\omega_2}  \Big] \nn\,,
\end{eqnarray}
where $h_v$ and $h_{v'}$ are Heavy Quark Effective Theory (HQET) fields for the
bottom and charm quarks, the index $j=L,R$ refers to the Dirac structures
$\Gamma ^h_{L}=\nslash P_L$ or $\Gamma ^h_{R}=\nslash P_R$, $\Gamma_c=({\bnslash
  P_L})/{2}$, $\xi_n^{(q)}$ are collinear quark fields and we sum over the
$q=u,d$ flavors. Note that no collinear strange quarks appear.  In
Eq.~(\ref{OV2}) the factors of $W$ and $S$ are Wilson lines required for gauge
invariance and the momenta subscripts $(\cdots)_{\omega_{i}}$ and
$(\cdots)_{k^+_{i}}$ refer to the momentum carried by the product of fields in
the brackets.  The matrix element of the soft fields in $O_L^{(0,8)}$ gives the
$S_L^{(0,8)}(k_1^+,k_2^+)$ distribution functions, for example
\begin{eqnarray}\label{Sintro}
 \frac{\langle D^{(*)0}(v') | (\bar h_{v'}^{(c)}S) \nslash P_{L} 
 (S^\dagger h_v^{(b)})
 (\bar d S)_{k^+_1}\nslash P_L (S^\dagger u)_{k^+_2} 
 | \bar B^0(v)\rangle}{\sqrt{m_B m_D}}
 &=& A^{D^{(*)}}\ S_{L}^{(0)}(k^+_1,k^+_2) \,,
\end{eqnarray}
where $A^D=1$ and $A^{D^*}=n\mcdot \varepsilon^*/n\mcdot v'=1$ (since the
polarization is longitudinal). The matrix element
of the collinear operator gives the LO light-cone distribution functions. We
work in the isospin limit and use the $(u\bar u+d\bar d)$, $s\bar s$ basis for
our quark operators. For $M=\eta,\eta'$ we have
\begin{eqnarray} \label{phi1}
\langle M(p) | \sum_{q=u,d} (\bar \xi_n^{(q)} W)_{\omega_1} 
   \frac{\bnslash\gamma_5}{\sqrt{2}}
  (W^\dagger \xi_n^{(q)})_{\omega_2}| 0\rangle
  &=& -i\, \bn\mcdot p\: f^{M}_q \:  \phi^{M}_q(\mu,x)\: \,,\\
\langle M(p) |(\bar \xi_n^{(s)} W)_{\omega_1} {\bnslash\gamma_5}
  (W^\dagger \xi_n^{(s)})_{\omega_2}| 0\rangle
  &=& -i\, \bn\mcdot p\: f^{M}_s \:  \phi^{M}_s(\mu,x)\: \,, \nn
\end{eqnarray}
while for vector mesons $M=\omega,\phi$ we simplify the dependence on the
polarization using $m_V\, \bn\mcdot \varepsilon^* = \bn\mcdot p$ and then have
\begin{eqnarray} \label{phi2}
\langle M(p,\varepsilon^*) | \sum_{q=u,d} (\bar \xi_n^{(q)} W)_{\omega_1} 
   \frac{\bnslash}{\sqrt{2}}
  (W^\dagger \xi_n^{(q)})_{\omega_2}| 0\rangle
  &=& i\, \bn\mcdot p\: f^{M}_q \:  \phi^{M}_q(\mu,x)\: \,,\\
\langle M(p,\varepsilon^*) |(\bar \xi_n^{(s)} W)_{\omega_1} {\bnslash}
  (W^\dagger \xi_n^{(s)})_{\omega_2}| 0\rangle
  &=& i\, \bn\mcdot p\: f^{M}_s \:  \phi^{M}_s(\mu,x)\: \,. \nn
\end{eqnarray}
In both Eq.~(\ref{phi1}) and (\ref{phi2}) we have suppressed a prefactor for
the $\phi^{M}$'s on the RHS:
\begin{eqnarray} \label{prefactor}
  \int_0^1\!\! dx\:
  \delta(\omega_1-x\, \bn\mcdot p)\, \delta(\omega_2+(1\!-\!x)\bn\mcdot p)\,.
\end{eqnarray}
Note that these definitions make no assumption about $\eta$-$\eta'$ or
$\omega$-$\phi$ mixing.  The SCET operators in Eq.~(\ref{OV2}) only give rise to
the $\phi^{M}_q$ terms. By charge conjugation $\phi^{M}_q(1-x)=\phi^{M}_q(x)$
and $\phi^{M}_s(1-x)=\phi^{M}_s(x)$ for both the isosinglet pseudoscalars and
isosinglet vectors. Our definitions agree with those in
Ref.~\cite{Kroll:2002nt}.

Now consider the graph emitting collinear gluons, Fig.~\ref{fig_scet1}(c).  and
integrate out the hard-collinear quark propagators to match onto
Fig.~\ref{fig_scet1}(h). Writing the result of computing this Feynman diagram in
terms of an operator gives a factor of $[ \bar h_{v'}^{(c)} \Gamma_j^h \{1,T^c\}
h_v^{(b)}]$ times
\begin{eqnarray} \label{step1}
 \big[ \bar d\, T^a \gamma_\perp^\mu P_L \{1,T^c\}
  \frac{\nslash}{2} \gamma_\perp^\nu T^b\, u\big]
 (ig{\cal B}_\perp^{\mu\,a})(ig{\cal B}_\perp^{\nu\,b}) \
 \frac{-\bn\mcdot p_2}{-\bn\mcdot p_2 n\mcdot k_2+i\epsilon} \:
 \frac{\bn\mcdot p_1}{\bn\mcdot p_1 n\mcdot k_1+i\epsilon} \, ,
\end{eqnarray}
where 
$ ig{\cal B}_{\perp\omega}^{\mu\,b}T^b = 
\big[{1}/{\bnP}\: W^\dagger [i\bn\mcdot D_c \,, iD_{c\perp}^\mu] W\big]_\omega $
is a LO gauge invariant combination with the gluon field strength.  The Dirac
structure can be simplified: $\gamma_\perp^\mu P_L \nslash \gamma_\perp^\nu =
-\nslash P_L (g_\perp^{\mu\nu}\!+\!  i\epsilon_\perp^{\mu\nu})$ where
$\epsilon^\perp_{12}=+1$. Furthermore we only need to keep operators that are
collinear color singlets, since others give vanishing contributions at this
order. These simplifications hold at any order in perturbation theory in \SCETa,
so the matching gives only two \SCETb operators
\begin{eqnarray} \label{GII}
 G_{j}^{(0)}(k^+_i,\omega_k) &=& 
  \Big[ \bar h_{v'}^{(c)}  \Gamma^h_j\,  h_v^{(b)} \:
  (\bar d\,S)_{k^+_1} \nslash P_L \, (S^\dagger u)_{k^+_2} \Big]
  \Big[ (g^\perp_{\mu\nu}\!+\!i\epsilon^\perp_{\mu\nu}) \,
    {\cal B}^{\mu\,b }_{\perp\omega_1} {\cal B}^{\nu\,b}_{\perp\omega_2} 
  \Big]\,, \\
 G_{j}^{(8)}(k^+_i,\omega_k) &=& 
   \Big[ \bar h_{v'}^{(c)}  \Gamma^h_jT^a\,  h_v^{(b)} \:
  (\bar d\,S)_{k^+_1} \nslash P_L T^a(S^\dagger u)_{k^+_2} \Big]
  \Big[ (g^\perp_{\mu\nu}\!+\!i\epsilon^\perp_{\mu\nu}) \,
     {\cal B}^{\mu\,b }_{\perp\omega_1} {\cal B}^{\nu\,b}_{\perp\omega_2} 
  \Big]\nn\,.
\end{eqnarray}
The operators in
Eq.~(\ref{GII}) appear as products of soft and collinear fields allowing us to
factorize the amplitude into soft and collinear matrix elements. We immediately
notice that the soft fields in Eq.~(\ref{GII}) and Eq.~(\ref{OV2}) are
identical.  Thus, the same non-perturbative $B\to D^{(*)}$ distribution
functions $S_L^{(0,8)}$ occur in the factorization theorem for the gluon and
quark contributions (cf. Eq.~(\ref{ftheorem})).  The matrix elements of the
collinear fields give
\begin{eqnarray} \label{phi3}
  M =\eta,\eta':\quad
  \langle M(p) | i\epsilon^\perp_{\mu\nu} \,
     {\cal B}^{\mu\,b }_{\perp,-\omega_1} {\cal B}^{\nu\,b}_{\perp,\omega_2}  
    | 0\rangle
   &=& 
    \frac{i}{2}\: \sqrt{C_F} f^{M}_1 \ 
   \overline \phi_{M}^{\,g}(\mu,x)
   \: \,,\\[3pt]
  M =\phi,\omega:\quad\ \:
 \langle M(p) | g^\perp_{\mu\nu}
  {\cal B}^{\mu\,b }_{\perp,-\omega_1} {\cal B}^{\nu\,b}_{\perp,\omega_2}
    | 0\rangle
  &=& \frac{i}{2}\: \sqrt{C_F}f^{M}_1\ \overline \phi_{M}^{\,g}(\mu,x)
  \: \,, \nn
\end{eqnarray}
where 
\begin{eqnarray}
  \overline \phi^M_{\, g}(x,\mu)= \frac{\phi^M_g(x,\mu)}{x(1\!-\!x)}\,,
\end{eqnarray}
$C_F=(N_c^2-1)/(2N_c)=4/3$, and $f^{M}_1=\sqrt{2/3}\, f^{M}_q+\sqrt{1/3}\,
f^{M}_s$. (We again suppressed a prefactor on the RHS of Eq.~(\ref{phi3}) which
is given in Eq.~(\ref{prefactor}).) Our $\phi_g^{\eta}$ and $\phi_g^{\eta'}$ are
the same as the ones defined in Ref.~\cite{Kroll:2002nt}, where they were used
to analyze the $\gamma$-$\eta$ and $\gamma$-$\eta'$ form factors.  Charge
conjugation implies
\begin{eqnarray}
  \phi^{M}_{g}(1-x) = - \phi^{M}_g(x) \,.
\end{eqnarray}
At tree level  using Eq.~(\ref{step1}) to match onto the gluon operators
$G_j^{(0,8)}$ gives
\begin{eqnarray} \label{GIIjet}
 J_g^{(0)} \!&=&\! \frac{\pi\alpha_s(\mu_0)}
  {N_c (n\mcdot k_2\!-\!i\epsilon)(n\mcdot
   k_1\!+\! i\epsilon)}\,, \qquad 
 J_g^{(8)} \!=\! \frac{\pi\alpha_s(\mu_0)}
  {(-N_c^3\!+\!N_c) (n\mcdot k_2\!-\!i\epsilon)
  (n\mcdot k_1\!+\!i\epsilon)}\,, 
\end{eqnarray}
where more generally $J_{g}^{(0,8)}= J_{g}^{(0,8)}(z,x,k_1^+,k_2^+)$. Thus, the
jet functions are even under $x\to 1-x$ while the gluon distributions are odd,
and the convolution in Eq.~(\ref{ftheorem}) for $A_{\rm glue}^{(*)M}$ vanishes.
Thus, $A_{\rm glue}^{(*)M}$ starts at ${\cal
  O}\big[\alpha_s^2(\sqrt{E\Lambda})\big]$ from one-loop corrections to the
gluon jet function.

The remaining contributions to the amplitude come from the isosinglet component
of the long distance operators shown in Figs.~\ref{fig_scet1}(d,e). These
operators take the form of a T-ordered product in \SCETb
\begin{eqnarray} \label{Obar}
  \overline O_{j}^{(0,8)}(\omega_i,k^+\!, \omega,\mu) &=& 
  \int\!\! \mbox{d}^4x\: T\,
   {\cal Q}^{(0,8)}_{j}(\omega_i,x=0)\ i L^{(0,8)}(\omega, k^+ \!, x) \,.
\end{eqnarray}
where $L^{(0,8)}(\omega, k^+ \!, x)$~\cite{Mantry:2003uz} are four quark
operators in \SCETb denoted by the shaded boxes in Figs.~\ref{fig_scet1}(d,e).
The matrix element of these long distance operators give the contribution
$A_{long}^{(*)M}$ in Eq.~(\ref{ftheorem}) where the collinear and soft functions
$\Psi_M^{(0,8)}$ and $\Phi_{L}^{(0,8)}$ are defined as
\begin{eqnarray} \label{PsiPhi}
  && \Big\langle M^0(p_M,\epsilon_M) \Big|  
  \Big[ (\bar\xi_n^{(d)} W)_{\omega_1}
  \bnslash P_L (W^\dagger \xi_n^{(u)} )_{\omega_2} \Big](0_\perp) 
  \Big[ (\bar\xi_n^{(u)} W)_{\omega}
  \bnslash P_L (W^\dagger \xi_n^{(d)} )_{\omega} \Big](x_\perp) 
  \Big| 0 \Big\rangle \nn\\
  &&\qquad 
   = i f^M/\sqrt{2}\:
   \Psi_M^{(0)}(z,\omega,x_\perp,\varepsilon_M^*) 
   \nn \,,\\
 && \Big\langle D^{(*)0}(v',\epsilon_{D^*}) \Big|  
  \Big[ (\bar h_{v'}^{(c)} S)
   \nslash P_{L}^h (S^\dagger h_v^{(b)} ) \Big](0_\perp) 
   \Big[ (\bar d S)_{k^+}
  \nslash P_L (S^\dagger u )_{k^+} \Big](x_\perp) 
  \Big| \bar B^0 \Big\rangle \nn\\
 && \qquad 
  = \sqrt{m_B m_{D^{(*)}}}\ \Phi_{L}^{(0)}(k^+,x_\perp,\varepsilon_{D^*}^*)  
  \,,
\end{eqnarray} 
and at tree level the jet functions are $\overline
J^{(0)}(\omega k^+)=-4/3\, \overline J^{(8)}(\omega k^+)=-8\pi\alpha_s(\mu)/(9
\omega k^+)$.

Eqs.~(\ref{Sintro},\ref{phi1},\ref{phi3},\ref{PsiPhi}) combined with
Eq.~(\ref{ftheorem}) completely define the amplitude for color suppressed decays
to leading nonvanishing order in $\Lambda _{\rm QCD}/Q$. We are now in a
position to make phenomenological predictions. We will neglect perturbative
corrections at the hard scale, $\alpha_s(Q)$. For heavy quark symmetry
predictions we will work to all orders in $\alpha_s(\sqrt{E\Lambda})$, while for
relating the $\eta$ and $\eta'$ amplitudes we will work to leading order in
$\alpha_s(\sqrt{E\Lambda})$.

The first class of predictions that we address make use of heavy quark symmetry
to relate the $D$ and $D^*$ amplitudes. It is worth mentioning why such
predictions are impossible to make using only HQET even though the $D, D^{*}$
are in a symmetry multiplet. If we do not factorize the energetic pion out of
the matrix element then the chromomagnetic operator which breaks the spin
symmetry comes in with a factor of $E_\pi/m_c\simeq 1.5$ and is not
suppressed~\cite{Mantry:2004qg}.  In the SCET analysis spin-symmetry breaking
effects are guaranteed to be suppressed by $\Lambda_{\rm QCD}/m_c$ allowing for
possible corrections at the $\sim 25\%$ level.

\begin{figure}[!t]
\vskip0.1cm
 \centerline{
  \mbox{\epsfxsize=12.truecm \hbox{\epsfbox{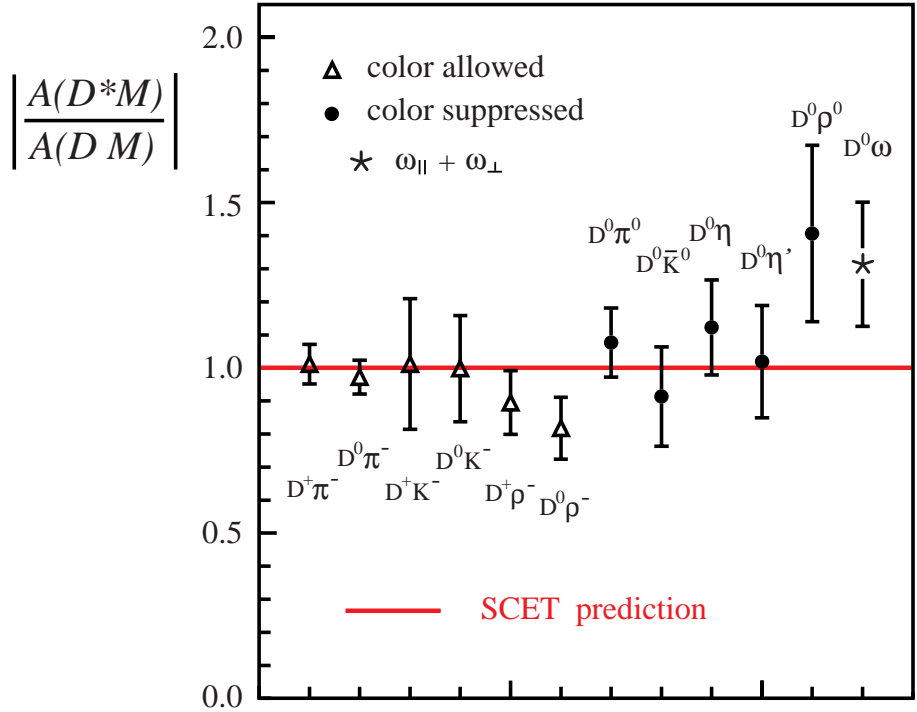}} }
  } 
 \vspace{0.2cm}
\vskip-0.3cm
\caption[1]{Comparison of the absolute value of the ratio of the amplitude
  for $B\to D^*M$ divided by the amplitude for $B\to DM$ versus data from
  different channels. This ratio of amplitudes is predicted to be one at leading
  order in SCET.  For $\omega$'s this prediction only holds for the longitudinal
  component, and the data shown is for longitudinal plus transverse.  }
\label{fig_data} 
\vskip0cm
\end{figure}
The factorization theorem in SCET, Eq.~(\ref{ftheorem}), moves the energetic
light meson into a separate matrix element. This allows us to use the formalism
of HQET in the soft sector to relate the $\bar B\to D$ and $\bar B\to D^*$
matrix elements in Eqs.~(\ref{Sintro}) and (\ref{PsiPhi}).  For $A_{short}^M$,
the contribution is the same for the $D$ and $D^*$ channels with identical soft
functions $S^{(i)}_L$ as a consequence of heavy quark symmetry.  The same is
true for the soft matrix element in $A_{\rm glue}$ which also gives $S^{(i)}_L$.
For the long distance contribution $A_{long}^M$, in addition to a dependence on
powers of $x_\perp ^2$, the soft function
$\Phi_{L}^{(i)}(k^+,x_\perp,\varepsilon_{D^*}^*)$ can have terms proportional to
$x_{\perp}\cdot \epsilon ^*_{D^*}$ in the $D^*$ channel while the collinear
function $\Psi_{M}^{(i)}(z,\omega ,x_\perp,\varepsilon_{M}^*)$ can have terms
proportional to $x_{\perp}\cdot \epsilon ^*_{M}$ in the case of vector mesons.
In the convolution over $x_\perp$ in $A_{long}^M$, the term in the integrand
proportional to the product $(x_{\perp}\cdot \epsilon ^*_{D^*}) (x_{\perp}\cdot
\epsilon ^*_{M})$ can be non-vanishing in the $D^*$ channel with a vector meson.
Such terms do not appear in the $D$ channel making the $D$ and $D^*$ amplitudes
unrelated in general.  However, if we restrict ourselves to longitudinal
polarizations, such terms in the $D^*$ channel vanish and the long distance
contributions in the two channels become identical. Finally, note that the
\SCETa jet functions, and the other collinear matrix elements in \SCETb are
identical for the two channels. Thus, at leading order in $\alpha _s{(Q)}$ and
$\Lambda _{\rm QCD}/Q$ the $D$ and $D^*$ channels are related as
\begin{eqnarray}
 && \frac{Br(\bar B\to D^*\eta)}{Br(\bar B\to D\eta)} = 
  \frac{Br(\bar B\to D^*\eta')}{Br(\bar B\to D\eta')} =
  \frac{Br(\bar B\to D^*\omega_\parallel)}{Br(\bar B\to D\omega)} 
  = 1 \ .
\end{eqnarray}
For the decay to $\phi$'s we also have
\begin{eqnarray}
 \frac{Br(\bar B\to D^*\phi_\parallel)}{Br(\bar B\to D\phi)} = 1 \,,
\end{eqnarray}
however in this case the prediction assumes that the
$\alpha_s^2(\sqrt{E\Lambda})$ contribution from $A_{\rm glue}$ dominates over
power corrections.  Note that we are expanding in $m_{M}/E_{M}$ so one might
expect the predictions to get worse for heavier states. For the case of color
suppressed decays to light mesons that are not isosinglets an analogous result
was obtained in Ref.~\cite{Mantry:2003uz}. It was shown that the long distance
contribution vanishes for $M=\pi, \rho $, so no restriction to longitudinal
polarization is required for $M=\rho$, but a restriction is needed for $M=K^*$.
Thus, for these color suppressed decays SCET predicts
\begin{eqnarray}
 &&\!\!\!\!
 \frac{Br(\bar B\to D^{*}\pi^0)}{Br(\bar B\to D\pi^0)} = 
  \frac{Br(\bar B\to D^{*}\rho^0)}{Br(\bar B\to D\rho^0)} =
  \frac{Br(\bar B\to D^{*}\bar K^0)}{Br(\bar B\to D \bar K^0)} =
  \frac{Br(\bar B\to D^{*}\bar K^{*0}_\parallel)}
      {Br(\bar B\to D \bar K^{*0})} 
   = 1 \,, \nn\\
 &&\!\!\!\!
 \frac{Br(\bar B\to D_s^{*+} K^-)}{Br(\bar B\to D_s^+ K^-)}
  = \frac{Br(\bar B\to D_s^{*+} K^{*-}_\parallel)}
     {Br(\bar B\to D_s^+ K^{*-})}=1 \,.
\end{eqnarray}
The factorization proven with SCET for color allowed decays~\cite{Bauer:2001cu}
also predicts the equality of the $D$ and $D^*$ branching fractions
~\cite{Politzer:1991au}.

Fig.~(\ref{fig_data}) summarizes the heavy quark symmetry predictions for cases
where data is available. We show the ratio of amplitudes because our power
expansion was for the amplitudes making it easier to estimate the uncertainty.
There is remarkable agreement in the color allowed channel where the error bars
are smaller and good agreement in the color suppressed channels as well.

So far our parameterization of the mixing between isosinglets in the
factorization theorem has been kept completely general, and we have not used the
known experimental mixing properties of $\eta$-$\eta'$ and $\phi$-$\omega$.  For
the next set of predictions we use the flavor structure of the \SCETb operators and
the isosinglet mixing properties to a) relate the $\eta $ and $\eta '$
channels and b) show that decays to $\phi$'s are suppressed.  Our discussion of
mixing parameters follows that in
Refs.~\cite{Leutwyler:1997yr,Feldmann:1997vc,Kaiser:1998ds,Feldmann:1998vh}.  In
general for a given isospin symmetric basis there are two light quark operators
and two states (say $\eta$ and $\eta'$) so there are four independent decay
constants. These can be traded for two decay constants and two mixing angles.
In an SU(3) motivated singlet/octet operator basis, $\{(\bar u u \!+\! \bar d d
\!+ \! \bar s s)/\sqrt{3}, (\bar u u \!+\! \bar d d \!-\! 2\bar s
s)/\sqrt{6}\}$, we have
\begin{eqnarray} \label{mix1}
 f^\eta_1= -f_1 \sin\theta_1\,,\quad  f^{\eta'}_1 = f_1 \cos\theta_1 \,,\quad
 f^\eta_8= f_8 \cos \theta_8 \,, \quad f^{\eta'}_8 = f_8 \sin\theta_8\,.
\end{eqnarray}
An alternative is the flavor basis used in Eq.~(\ref{ftheorem}) ,
$\{O_q,O_s\}\sim \{(\bar u u \!+\!  \bar d d)/\sqrt{2},\, \bar s s\}$. Here
\begin{eqnarray} \label{mix2}
  f^\eta_q= f_q \cos\theta_q \,, \quad 
  f^{\eta'}_q = f_q \sin\theta_q \,, \quad
 f^\eta_s = -f_s \sin\theta_s \,, \quad 
 f^{\eta'}_s = f_s \cos \theta_s \,.
\end{eqnarray}
Phenomenologically, $(\theta_8-\theta_1)/(\theta_8+\theta_1) \simeq 0.4$ which
can be attributed to sizeable SU(3) violating effects, whereas
$(\theta_q-\theta_s)/(\theta_q+\theta_s)\lesssim 0.06$ where a non-zero value
would be due to OZI violating effects~\cite{Feldmann:1999uf}. We therefore adopt
the FKS mixing scheme~\cite{Feldmann:1998vh,Feldmann:1999uf} where OZI violating
effects are neglected and the mixing is solely due to the anomaly. Here one
finds experimentally
\begin{eqnarray} \label{FKS}
 \theta_q\simeq \theta_s\simeq \theta= 39.3^\circ \pm 1.0^\circ \,.
\end{eqnarray}
Thus it is useful to introduce the approximately orthogonal linear combinations 
\begin{eqnarray} \label{mixstates}
  |\eta_q \rangle = \cos\theta\: |\eta\rangle +  \sin\theta\: |\eta'\rangle \,,
    \qquad
 |\eta_s\rangle = -\sin\theta\: |\eta \rangle +  \cos\theta\: |\eta' \rangle \,,
\end{eqnarray} 
since neglecting OZI effects the offdiagonal terms $\langle 0 | O_q |
\eta_s\rangle$ and $\langle 0 | O_q | \eta_s\rangle$ are zero. Since this is
true regardless of whether these operators are local or non-local, the matrix
elements in Eqs.~(\ref{phi1},\ref{PsiPhi}) must obey the same pattern of mixing
as in Eq.~(\ref{mix2}) [$f^\eta_q \phi^\eta_q(x) = f_q\phi_q(x) \:\cos\theta_q,$
etc.]  and so
\begin{eqnarray} \label{mix3}
 \phi_q^\eta(x) = \phi_q^{\eta'}(x) = \phi_q(x) \,,\qquad 
 \phi_s^\eta(x) = \phi_s^{\eta'}(x) = \phi_s(x) \,, \qquad
 \Psi_\eta^{(0,8)}= \Psi_{\eta'}^{(0,8)}= \Psi_q^{(0,8)}\,.
\end{eqnarray}

The \SCETb operators of Eq.~(\ref{GII}) which contribute to $A_{glue}^{(*)M}$
can produce both the $\eta _q$ and $\eta _s$ components of the isosinglet
mesons. However, recall that at LO in $\alpha_s(\sqrt{E\Lambda})$ the
convolution over the momentum fractions in $A_{glue}^{(*)M}$ vanishes allowing
us to ignore this contribution.  The remaining contributions from
$A_{short}^{(*)M}$ and $A_{long}^{(*)M}$ involve operators that can only produce
the $\eta _q$ component of the isosinglet mesons as seen by the flavor structure
of the operators in Eqs.~(\ref{OV2}) and (\ref{PsiPhi}). We can now write the
amplitude for the $\eta ^{(')}$ channels in the form
\begin{eqnarray}
A^{(*)\eta} = \cos\theta\: 
  [A^{(*)\eta _q}_{\rm short} +A^{(*)\eta _q}_{\rm long} ]\,,\qquad
A^{(*)\eta '} = \sin\theta\: 
  [A^{(*)\eta _q}_{\rm short} +A^{(*)\eta _q}_{\rm long} ].
\end{eqnarray}
This leads to a prediction for the relative rates with SCET
\begin{eqnarray} \label{mixpredict}
 \frac {Br(\bar B\to D\eta ')}{Br(\bar B\to D\eta )} =
 \frac {Br(\bar B\to D^{*}\eta ')}{Br(\bar B\to D^{*}\eta )} =
  \tan^2(\theta)=0.67 \,,
\end{eqnarray}
with uncertainties from $\alpha_s(\sqrt{E\Lambda})$ that could be at the $\sim
35\%$ level.  Experimentally the results in Table~\ref{table_data} imply
\begin{eqnarray} \label{ratio}
  \frac {Br(\bar B\to D\eta ')}{Br(\bar B\to D\eta )} = 0.61 \pm 0.12\,,\qquad
  \frac {Br(\bar B\to D^{*}\eta ')}{Br(\bar B\to D^{*}\eta )} = 0.51 \pm 0.18 \,,
\end{eqnarray}
which agree with Eq.~(\ref{mixpredict}) within the $1$-$\sigma$ uncertainties.

For the isosinglet vector mesons we adopt maximal mixing which is a very good
approximation (meaning minimal mixing in the FKS basis), and is consistent with
the anomaly having a minimal effect on these states and with neglecting OZI
effects. In this case only $\langle 0 | O_q | \omega\rangle$ and $\langle 0 |
O_s | \phi \rangle $ are non-zero. Thus only $A_{short}^{(*)\omega}$ and
$A_{long}^{(*)\omega}$ are non-zero and we predict that $\phi $ production is
suppressed
\begin{eqnarray}
 \frac{Br(\bar B^0\to D^{(*)0}\phi)}{Br(\bar B^0\to D^{(*)0}\omega)}
   = {\cal O}\Big(\alpha_s^2(\sqrt{E\Lambda}), \alpha_s(\sqrt{E\Lambda})\frac{\Lambda_{\rm QCD}}{Q},
     \frac{\Lambda_{\rm QCD}^2}{Q^2} \Big) \lesssim 0.2 \,,
\end{eqnarray}
possibly explaining why it has not yet been observed. Interestingly a
measurement of $\bar B\to D\phi$ or $\bar B\to D^*\phi$ may give us a direct
handle on the size of these expansion parameters.


Just using the original form of the electroweak Hamiltonian in Eq.~(\ref{Hw})
there is an SU(3) flavor symmetry relation among the color suppressed
decays~\cite{Savage:1989ub}
\begin{eqnarray} \label{RR}
  R_{\rm SU(3)} &=& 
    \frac{ Br(\bar B^0\to D_s^+ K^-)}{Br(\bar B\to D^0\pi^0)} 
   + \bigg|\frac{V_{ud}}{V_{us}}\bigg|^2 
     \frac{ Br(\bar B^0\to D^0 \bar K^0)}{Br(\bar B\to D^0\pi^0)} 
   - \frac{3 Br(\bar B^0\to D^0 \eta_8)}{Br(\bar B\to D^0\pi^0)} = 1 \,,\\
  R^*_{\rm SU(3)} &=& 
   \frac{ Br(\bar B^0\to D_s^{*+} K^-)}{Br(\bar B\to D^{*0}\pi^0)} 
   +  \bigg|\frac{V_{ud}}{V_{us}}\bigg|^2 
     \frac{ Br(\bar B^0\to D^{*0} \bar K^0)}{Br(\bar B\to D^{*0}\pi^0)} 
   - \frac{3 Br(\bar B^0\to D^{*0} \eta_8)}{Br(\bar B\to D^{*0}\pi^0)} = 1 \,,\nn
\end{eqnarray}
where $\eta_8$ is the SU(3) octet component of the $\eta$. In the SU(3) limit
the $\eta-\eta'$ mixing vanishes and we can take $\eta_8=\eta$. Away from this
limit there is SU(3) violation from the mixing as well as from other sources,
and it is the latter that we would like to study.  To get an idea about the
effect of mixing we set $|\eta_8\rangle = \cos\vartheta |\eta\rangle
+\sin\vartheta | \eta' \rangle$, which from Eq.~(\ref{mixstates}) can then be
written in terms of $|\eta_q\rangle$ and $|\eta_s\rangle$, and vary $\vartheta$
between $-10^\circ$ and $-23^\circ$.  From the flavor structure of the leading
order SCET operators for $B\to DM$ decays we then find
\begin{eqnarray} 
  \frac{Br(\bar B^0\to D\eta_8)}{Br(\bar B^0\to D\eta)} &=& 
    \frac{Br(\bar B^0\to D^*\eta_8)}{Br(\bar B^0\to D^*\eta)} = 
    \frac{\cos^2(\theta-\vartheta)}{\cos^2(\theta)} \,,
\end{eqnarray}
where $\vartheta$ is the $\eta$-$\eta'$ state mixing angle in the flavor
octet-singlet basis and $\theta$ is the FKS mixing angle. In the SU(3) limit
$\vartheta=\theta_1=\theta_8=0$, however phenomenologically $\vartheta\simeq
-10^\circ$ to $-23^\circ$.  Experimentally taking $|V_{us}/V_{ud}|=0.226$ and
using Table~\ref{table_data} gives
\begin{eqnarray} \label{RR2}
 R_{\rm SU(3)} &=& \Bigg\{ \begin{array}{l} 1.00 \pm 0.59  \ \ [\vartheta = 0^\circ] \\
   1.75 \pm 0.57 \ \ [\vartheta = -10^\circ] \\ 
   2.64 \pm 0.56 \ \ [\vartheta = -23^\circ] \end{array} 
   \,,\quad 
 R^*_{\rm SU(3)} =\Bigg\{ \begin{array}{l} -0.22 \pm 0.97 \ \ [\vartheta = 0^\circ]\\
  \phantom{-}0.59 \pm 0.88  \ \ [\vartheta = -10^\circ]\\ 
  \phantom{-}1.57 \pm 0.83 \ \ [\vartheta = -23^\circ] \end{array}
  \,. \ \ 
\end{eqnarray}
In all but one case the central values indicate large SU(3) violation, however
the experimental uncertainty is still large. It would be interesting to compute
the uncertainties by properly accounting for correlations between the data
rather than assuming these correlations are zero as we have done. At
$1$-$\sigma$ the errors accommodate $R^*_{\rm SU(3)}=1$ except if
$\vartheta=0^\circ$, and only accommodate $R_{\rm SU(3)}=1$ if
$\vartheta=0^\circ$. Note that the heavy quark symmetry prediction, $R^*_{\rm
  SU(3)}=R_{\rm SU(3)}$, is still accommodated within the error bars.

In the pQCD approach predictions for color suppressed decays to isosinglets have
been given in Refs.~\cite{Keum:2003js,Lu:2003xc}, where they treat the charm as
light and expand in $m_c/m_b$. With such an expansion there is no reason to
expect simple relationships between decays to $D$ and $D^*$ mesons because heavy
quark symmetry requires a heavy charm. In Ref.~\cite{Lu:2003xc} predictions for
$\eta$ and $\eta'$ were given dropping possible gluon contributions.  Our
analysis shows that this is justified and predicts a simple relationship between
these decays, given above in Eq.~(\ref{mixpredict}).

To conclude, we derived a factorization theorem which describes color suppressed
decays to isosinglets solely from QCD without model dependent assumptions by
expanding in $\Lambda_{\rm QCD}/Q$. Phenomenological implications were discussed
for $B\to D\eta$, $D\eta'$, $D\omega$, $D\phi$.  We proved that the gluon
production amplitude involves the same soft $B\to D$ matrix element as the
non-gluon terms.  We then showed that the factorized form of the amplitudes
together with heavy quark symmetry predict that ${Br(\bar B\to D^*\{\eta, \eta
  ',\omega _{||},\phi _{||}\})} = {Br(\bar B\to D\{\eta, \eta
  ',\omega,\phi\})}$, with corrections being suppressed by either a power
$\Lambda_{\rm QCD}/Q$ or a factor of $\alpha_s(Q)$. The $\alpha_s(Q)$ terms can
be computed in the future.  We also consider $\eta$-$\eta'$ mixing and showed
that due to the vanishing of the gluon contributions the flavor structure of the
SCET operators imply ${Br(\bar B\to D^{(*)}\eta ')}/{Br(\bar B\to D^{(*)}\eta )}
= \tan^2(\theta) = 0.67 $ where $\theta =39.3^\circ$ is the $\eta -\eta '$
mixing angle in the FKS scheme, and that $Br(\bar B\to D^{(*)}\phi)/Br(\bar B\to
D^{(*)}\omega)\lesssim 0.2$.  Corrections here are only order
$\alpha_s(\sqrt{E\Lambda})$ and should be computed in the near future. At
one-loop the effect of operator mixing will also need to be
considered~\cite{Fleming:2004rk}.  Finally, tests of SU(3) symmetry were given
in Eqs.~(\ref{RR}-\ref{RR2}).

\acknowledgments A.B. and S.M. would like to thank the organizers of TASI 2004
where part of this work took place. This work was supported in part by the U.S.
Department of Energy (DOE) under the cooperative research agreement
DF-FC02-94ER40818. I.S.  was also supported in part by a DOE Outstanding Junior
Investigator award and an Alfred P.~Sloan Fellowship.

\bibliography{bdeta}

\end{document}